\documentclass[twocolumn,showpacs,preprintnumbers,amsmath,amssymb]{revtex4}

\usepackage{dcolumn}
\usepackage{bm}
\usepackage{subfigure}
\usepackage{amsmath}
\usepackage{amsfonts}   
\usepackage{graphics}  
\usepackage{psfrag} 
\usepackage{graphicx} 
\usepackage{epsfig}    
\usepackage{rotating}  
\usepackage[latin1]{inputenc}
\usepackage{color}
\usepackage{rotating}
\usepackage{lscape}
\usepackage[bookmarks=false]{hyperref}

\newcommand{\ru}{\rho^{\uparrow}}
\newcommand{\rd}{\rho^{\downarrow}}
\newcommand{\ud}{{\uparrow\left(\downarrow\right)}}

\begin{document}

\preprint{}

\title{Driven Transport on Parallel Lanes with Particle Exclusion and Obstruction}

\author{Anna Melbinger$^1$, Tobias Reichenbach$^2$, Thomas Franosch$^{1,3}$ and Erwin Frey$^1$}
\affiliation{$^1$Arnold Sommerfeld Center for Theoretical Physics (ASC) and
Center for NanoScience (CeNS), Department of Physics,
Ludwig-Maximilians-Universit\"at M\"unchen,
Theresienstra\ss e 37, D-80333 M\"unchen, Germany\\
$^2$Laboratory of Sensory Neuroscience, The Rockefeller University,
1230 York Avenue, New York, NY 10065, U.S.A.\\
$^3$ Institut f\"ur Theoretische Physik,  Universit\"at Erlangen-N\"urnberg,  
Staudtstra{\ss}e 7,D-91058 Erlangen, Germany
}

\date{\today}

\begin{abstract}
We investigate a  driven two-channel system where particles on different lanes mutually
obstruct each others motion extending an earlier model by Popkov and Peschel~\cite{Peschel}. This obstruction may occur in biological contexts due to steric hinderance  where
motor proteins carry cargos by `walking' on microtubules. Similarly, the model serves as a description for classical spin
transport where charged particles with internal states move unidirectionally on a lattice.
Three regimes of qualitatively different behavior are identified depending on the strength of coupling between the lanes.
For small and large coupling strengths
the model can be mapped to a one-channel problem, whereas a new and rich phase behavior emerges for intermediate ones.
We derive an approximate but quantitatively accurate theoretical description in terms of a one-site cluster approximation, 
and obtain insight into the phase behavior through the  current-density relations combined with  an extremal-current principle. 
Our results are confirmed by stochastic simulations. 
\end{abstract}

\maketitle

\section{Introduction}\label{sec_Intro}
Driven diffusive systems  are of importance in various fields of physics and biology~\cite{SchmittmannZia, schuetz_book}, 
since they  serve as simplistic models for biological transport phenomena~\cite{hirokawa-1998-279, chou-1999-82, Howard, Hinsch}, 
traffic flow~\cite{Peschel,helbing-2001-73, chowdhury-2000-329,schmittmann-2005-70}, fast ionic conductors~\cite{PhysRevB.28.1655}, as well
as quasi-classical spin-transport~\cite{reichenbach-2006-97}. Furthermore  they  provide valuable
insights into non-equilibrium statistical mechanics. As an example, and in
contrast to equilibrium systems, their bulk behavior is sensitive to the boundaries~\cite{krug-1991-76}. Boundary induced phase transitions  in one dimension may emerge resulting
in complex phase behavior.

The most prominent example of driven lattice gases, the Totally Asymmetric Exclusion Process (TASEP), was originally
proposed as a simple model for 
the motion of multiple ribosomes along a mRNA strand during protein translation~\cite{macdonald-1968-6}. In this model
particles move unidirectionally along a one-dimensional lattice, provided the next site is empty.  Exact solutions, \emph{e.g.} by
employing the Bethe ansatz~\cite{Derrida-1998-80} or a matrix product ansatz~\cite{derrida-1993-26}, are feasible, yet
much insight can readily be obtained from simple mean-field considerations~\cite{schuetz_book}.

Intracellular transport  constitutes another fascinating biological application~\cite{vilfan-2001} of driven systems. Here, 
molecular motors such as kinesin or dynein, driven by the hydrolysis of  adenosine triphosphate (ATP), move
unidirectionally along microtubules~\cite{hirokawa-1998-279}.  Macromolecules or other cellular constituents, which often
are too large to diffuse fast enough through the crowded cytosol, are carried by motor proteins, and are then actively transported
to the location where they are needed. Recent theoretical studies motivated  by these processes have investigated the influence
of attachment and detachment of the motors to the microtubules~\cite{parmeggiani-2003-90,klumpp-2003-113,parmeggiani-2004,nishinari-2005-95,greulich-2007-75}, 
extended particles~\cite{pierobon-2006-74}, the influence of defects on the track~\cite{PhysRevLett.78.3039,pierobon-2006-74-2,greulich-2008}, and the
competition between different motor species~\cite{ComertKural06032005,muller-2008-105}.  Further attention has been paid to transport along
several coupled channels where particles move in parallel. This coupling can be either achieved by allowing lane-switching
events~\cite{knospe-2002-35,reichenbach-2006-97,reichenbach-2007-9,refId,pronina-2004-37,jiang:041128,mitsudo-2005-38,juhasz-2007-76, chowdhury:050902, ISI:000260132000001} or by a possible influence of a particle  in one channel on the motion in the other channel~\cite{pronina-2007-40, Peschel}. 
Here we consider the latter case and investigate how mutual obstruction of motor proteins on neighboring lanes, for example
stemming from large cargos attached to them, affects the transport properties of the system.

Driven diffusive systems may also serve as a description for spin transport with possible applications  in the field of spintronics. 
For instance, such spin currents flow in a chain of quantum dots where electrons are driven by an external voltage in a way 
that only the lowest energy levels can be occupied~\cite{hanson2007sfe}. Hence, not more than one electron of each internal
state is permitted per site and electrons located in the immediate vicinity repel each other due to Coulomb interaction. 
A model taking Pauli's exclusion principle into account while ignoring phase coherence has been investigated recently~\cite{reichenbach-2006-97,reichenbach-2007-9,refId}, yet Coulomb blockade has been neglected.  
In the present article we focus on the influence of a mutual obstruction mimicking for example Coulomb interaction. To identify its effect on
the collective transport properties in the clearest way, we disregard spin-flip events which can cause intriguing behavior on
their own~\cite{reichenbach-2006-97,reichenbach-2007-9,refId}.

A simple lattice model that incorporates mutual obstruction on two lanes has been investigated by Popkov and Peschel~\cite{Peschel}. The steric hindrance there is manifested in the hopping rates that explicitly depend on the configuration on the opposing lane. As a consequence of this coupling of the lanes 
a variety of peculiar phases arises which has been neatly rationalized in terms of a cluster approximation. Particularly, symmetry breaking, which arises even though the boundary conditions are symmetric, is observed and analyzed.

In this article we extend the model of Ref.~\cite{Peschel} by considering asymmetric boundary conditions and rates instead of reservoirs at the right boundary.
We introduce the model in Section \ref{sec_Model}, both in the two-lane and in the spin-transport picture. 
In Section~\ref{sec_Classification} we describe the stochastic simulations and provide first insights in how particle obstruction 
affects the behavior. Namly, we identify three regimes of qualitatively different behavior. In Section~\ref{CDR}, we analytically compute the current-density relations within a one-site cluster approximation. Section~\ref{PB} presents a discussion on how the current-density relation, obtained via an extremal-current principle, allows to identify the system's different phases and to analytically predict the phase diagram. We summarize our main findings in Section~\ref{summary} and provide a brief conclusion.

\section{Model}\label{sec_Model}

We examine a driven diffusive system  which serves as a minimal model for the transport on two parallel lanes which are coupled by a repulsive short-range interaction. The same model describes classical driven spin transport with  Coulomb blockade. In the following we specify the dynamics in detail,  presenting both the two-lane and the spin-transport representation.

\subsection{Two-channel Representation}

\begin{figure}[h]
\includegraphics[width=0.45\textwidth]{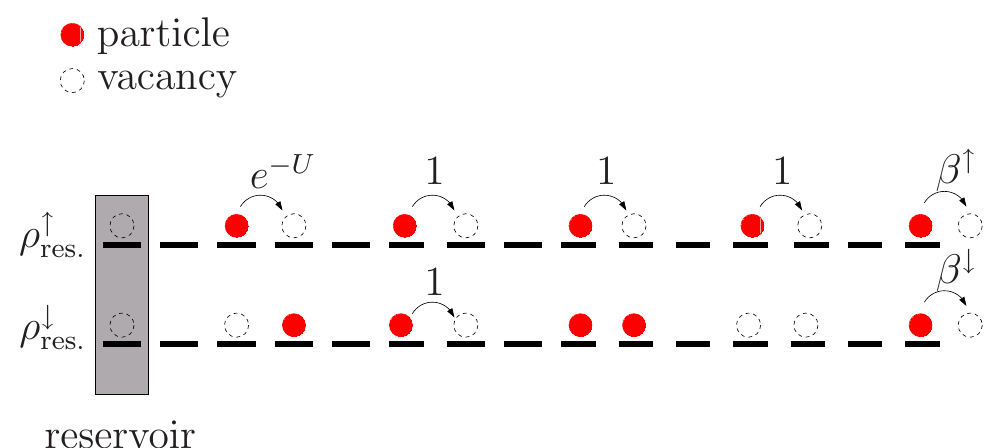}
\caption{(color online) Illustration of the two-lane representation. Particles enter from two reservoirs at the left boundary with 
densities $\ru_\text{res},~\rd_\text{res}$. In bulk, hopping rates depend on the particle configuration of the, respective, other lane. 
At the right boundary  particles leave at rates $\beta^\uparrow,~\beta^\downarrow$.}
\label{fig:2lane}
\end{figure}
Consider particle transport along two parallel channels, each of them containing $N$ discrete lattice sites, 
see Fig.~\ref{fig:2lane}. Each site may contain at most one particle (on-site exclusion), such that the occupation number
of site $i$ on the upper (lower) channel, $n_i^\uparrow~(n_i^\downarrow)$, can only take values $0$ or $1$, corresponding 
to a vacant or an occupied site, respectively.

Particles enter from two entangled reservoirs located at 
the left-hand side of the system. At each time step, the reservoir is in one of the four possible states: (i)~double occupation with 
probability $\kappa^*$, (ii)~only the upper reservoir is occupied with probability $\ru_\text{res}-\kappa^*$, (iii)~only the lower reservoir
is occupied with probability $\rd_\text{res}-\kappa^*$ and (iv)~both reservoir sites are empty with probability $1-\ru_\text{res}-\rd_\text{res}+\kappa^*$. 
Thus, $\ru_\text{res}$ and $\rd_\text{res}$ are the average densities on the upper and lower reservoir, respectively, and $\kappa^*$ 
corresponds to the double-occupation density in the reservoir.

In bulk, particles move unidirectionally to the right. Due to obstruction, 
the hopping rate thereby depends on the particle configuration at the other lane. A particle attempting to proceed by one site is obstructed if 
a particle resides on the subsequent site of the other channel. However, this obstruction is relevant only when the particle does not 
experience obstruction in its current position, meaning when its current neighboring site on the other channel is empty.  
We model this effect by reducing the hopping rate to a value $e^{-U}$ $(U>0)$ in this case, while the rate is unity for all other configurations. 
In the biological context of molecular motors walking along microtubuli, the reduced hopping rate corresponds to the spatial obstruction 
stemming from large cargos attached to motor proteins. 

Last, the rules of the model are completed by specifying how particles leave the system after traversing the bulk. Here, we consider that particles at the right boundary leave with the exiting rates $\beta^\uparrow,~\beta^\downarrow$ in contrast to ~\cite{Peschel}.

\subsection{Spin-Transport Representation}

\begin{figure}[h]
\includegraphics[width=0.45\textwidth]{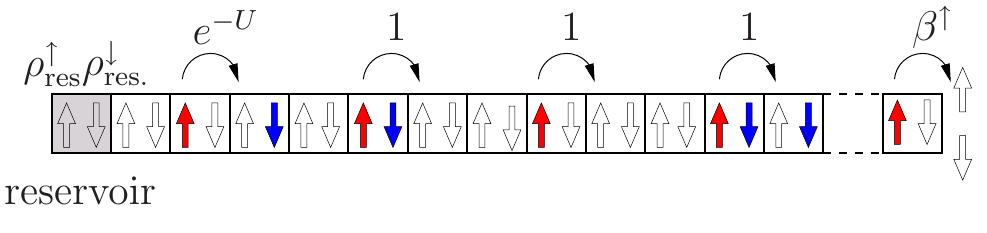}
\label{fig:spin-transport}
\caption{(color online) Illustration of an exclusion model with two internal states, adopting the language of spin transport. Particles in 
state $\uparrow$ ($\downarrow$) enter the system at the left boundary from a reservoir with density $\rho^\uparrow_\text{res.}~(\rho^\downarrow_\text{res.})$ 
and leave at the right boundary with rate $\beta^\uparrow~(\beta^\downarrow)$. In bulk particles hop to the right always respecting Pauli's exclusion principle. 
If an unpaired particle moves to a site which is already occupied by a particle of the other spin state the hopping rate is decreased 
to $e^{-U}$, otherwise, the hopping rate is set to 1. }
\end{figure}
The model can be readily interpreted in the context of spin transport where it  serves as a description 
for classical spin currents. 
The analogy to the two-channel picture is the following:  a particle situated at the upper (lower) lane maps to
a particle with spin up (spin down). At the left boundary particles enter from  a spin reservoir with densities $\ru_\text{res},~\rd_\text{res}$. 
Having traversed the lattice, they leave the system at the right boundary with exiting  rates $\beta^\uparrow,~\beta^\downarrow$. 
In bulk the particles move to the right always respecting Pauli's exclusion principle, \emph{i.e.} only one particle per internal state is 
permitted per site. According to the two-lane representation, the hopping rates depend on the particle configuration of the system. A  short-range 
repulsive interaction reduces the hopping rate for an unpaired particle onto a site which is already occupied by a particle of the
other spin state to $e^{-U}$, as compared to $1$ for the other configurations. The parameter $U>0$ may be viewed as an effective interaction potential, 
originating from a repulsive Coulomb interaction, where particles on the same site (though different spin states) gain potential energy. 
In this context, one can also consider an increased hopping rate away from a double occupation, yet, one can show that this does not change
our results qualitatively~\cite{Dipl-arbeit}.

For clarity, we employ  only the two-channel representation in the following.

A similar model as the one introduced above was recently proposed in~\cite{Peschel}. However, only symmetric situations where considered with entrance/exit reservoirs that were equal an both lanes. As a further difference to our model the authors did model the exiting processes through reservoirs at the right side instead of exiting rates. Because we explicitly investigate the asymmetric case, with entrance/exit properties that differ for the two lanes, and because of our usage of exit rates we find a multitude of new phenomena, summarized in Section~\ref{summary}. The asymmetry between the two lanes requires a two-dimensional generalization of the extremal-current principle. The derivation of this two-dimensional extremal-current principle constitutes a key result of our work; we show how it successfully describes much of the system's behavior.

\section{Classification of the Systems Sensitivity on the Potential}\label{sec_Classification}

The steady-state bulk densities $\ru_i=\langle n_i^\uparrow\rangle,~\rd_i=\langle n_i^\downarrow\rangle$, where $\langle\cdot\rangle$ indicates a coarse-grained time average, constitute key observables.
Because of particle conservation their temporal evolution  can be obtained from the particle flux $j_{i-1}$  onto 
site $i$ and the one away from it, $j_i$:
\begin{eqnarray}
\partial_t\ru_i&=&j^\uparrow_{i-1}-j^\uparrow_{i},\nonumber\\
\partial_t\rd_i&=&j^\downarrow_{i-1}-j^\downarrow_{i}.
\label{Eq:time_ev}
\end{eqnarray}
The currents, $j_i^\uparrow,~j_i^\downarrow$, contain correlations between neighboring sites on the lattice. To find an analytic description 
these correlations have to be accounted for by a suitable closure relation, \emph{e.g.} by a mean-field approximation or a one-site cluster approximation.

Stochastic simulations  provide another route to gain insight into the system's behavior. 
In this Section we first detail the simulation algorithm, and then describe  three classes of behavior that emerge for different interaction strength.

\subsection{Stochastic Simulations}

We have determined the system's stationary state via stochastic simulations with random sequential updating, using the dynamic rules introduced
in the previous section and employing the Gillespie algorithm~\cite{gillespie-1976-22, gillespie-1977-81}. We have performed  time averages 
over about $10^5$ time intervals, each containing  $10\times L$ time steps and the lattice size is set to $L=1000$.
At the left boundary, the reservoir dynamics is specified in terms of the three parameters, $\ru_\text{res},~\rd_\text{res}$ and $\kappa^*$. 
Here, we restrict the discussions to the case of \emph{relaxed} reservoirs, where the correlations in the reservoirs reflect the ones in bulk, which is particularly
illuminating and amenable to a theoretical description. Then, the double occupation density can be determined from the average
densities, $\ru_\text{res},~\rd_\text{res}$ according to Eq.~(\ref{eq:kappa}) derived in the Appendix.
In general, and apart from boundary effects such as boundary layers,  we found constant density profiles in the system. To determine the corresponding 
value of the average density in bulk for constant density profiles, we only considered the $0.2\times L$ sites in the center of both channels.
Our simulations confirm  to a large extent the analytic approximations which are to be discussed in the following sections.

\subsection{Dependence on the Interaction Strength $U$}

\begin{figure}[t]
\includegraphics[width=0.45\textwidth]{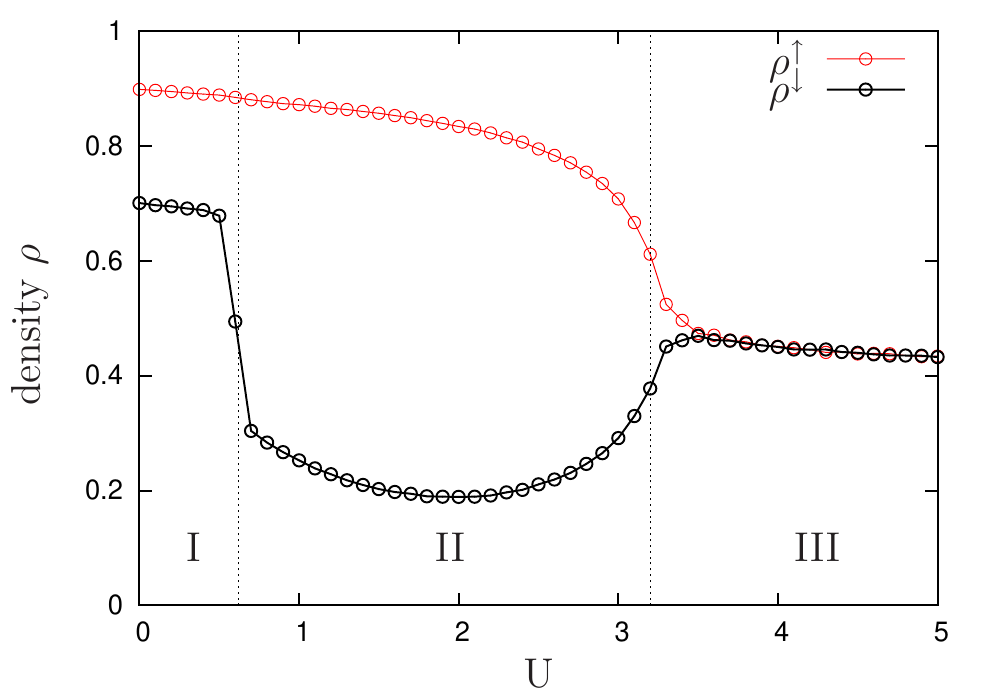}
\caption{(color online) Average bulk densities obtained by stochastic simulations on the upper (red, grey) and lower lane (black) for varying  potential strength $U$. 
The parameters are $\ru_\text{res}=\rd_\text{res}=0.5,~\beta^\uparrow=0.1$ and$~\beta^\downarrow=0.3$. Three regimes of qualitatively different
behavior emerge. In the first one (I) the system qualitatively behaves like two uncoupled systems. In regime II, the density in the lower
channel strongly decreases, while the density in the upper channel is still large.
This nontrivial behavior is discussed in detail 
in Section~\ref{CDR} and Section~\ref{PB}. In regime III, for strong coupling, the system behaves like a one-channel TASEP.}
\label{fig:dep_U}
\end{figure}

In the case of vanishing coupling, \emph{i.e.} $U=0$,  the system simply corresponds to two uncoupled TASEPs. 
In the presence of obstruction, and thereby coupling  between both channels, this picture changes drastically upon increasing the effective
interaction strength. Our stochastic simulations show three regimes of qualitatively different behavior, which are illustrated in  Fig.~\ref{fig:dep_U}. In the following, we discuss these regimes, and provide a mapping on TASEP for two of them.

\subsubsection{Weak Coupling}

In the first regime (I), for small coupling strength $U$,  the system almost behaves like two uncoupled lanes, except
for the fact that the densities are slightly reduced. This regime can be well described by a simple mean-field approximation, 
where correlations between different lattice sites, $i\neq j$, are neglected, $\langle n^\ud_i n^\ud_j\rangle\approx \langle n^\ud_i\rangle \langle n^\ud_j \rangle$, 
or by a one-site cluster approximation which is to be discussed in detail in the following section. In this regime the phase 
behavior qualitatively corresponds to the one already known from TASEP. On a quantitative level, differences arise as  the phase transition 
lines are shifted compared to the uncoupled case.

\subsubsection{Intermediate Coupling}

In regime II, the one of intermediate coupling strength, an intriguing new phase behavior emerges. For instance, for the set 
of parameters shown in Fig.~\ref{fig:dep_U}, the density in the upper channel remains rather undisturbed by the obstruction, while the one in the 
lower channel drops to a comparatively small value. The nontrivial behavior in regime II is caused by the influence of the potential on the transport 
properties in bulk as well as on the boundaries. Especially, the exiting current is strongly influenced by the interaction potential  
resulting in smaller densities at the right boundary than expected for uncoupled systems. We rationalize this behavior in the following section. 
Because the systems operates far from equilibrium, this change in the boundary conditions has a strong impact on the system.

Secondly, the transport properties in bulk also react sensitively to the coupling. As discussed in detail in the following section, this causes 
changes in the currents' dependences on the bulk densities for potentials larger than a critical strength $U_\text{C}=\ln 4\approx1.4$.

Furthermore,  other intriguing phenomena,  such as domain walls between a low and a high density phase, are found. Also, there exist phases where 
the system depends sensitively on both boundaries, namely, the total current of the system is fixed by the right boundary while the exact
value of the densities in bulk depends on the left boundary.

\subsubsection{Strong Coupling}

A further increase in $U$  leads to strong obstruction between the lanes with new qualitative behavior. In this regime of strong coupling (III) 
double occupancy of a site almost never occurs due to the vanishing hopping rate $e^{-U}$ that would yield this configuration. 
Therefore, the transport properties of two coupled  channels 
is similar to a single-channel TASEP.
This mapping cannot always be performed if the boundaries are reservoirs. It then only holds for the special case of small reservoir densities. Otherwise correlations between the upper and the lower reservoir are large. They are transported in the bulk and there destroy the effective one-lane behavior. Especially, the behavior shown in Fig.~\ref{fig:dep_U} would drastically change, if the exiting rates $\beta^\uparrow=0.1$ and $\beta^\downarrow=0.3$ would be replaced by the corresponding right reservoirs, namely the description holding for regime II would then apply and the bulk densities would not show the one-channel behavior.

Introducing the total density, $\tau_i=\langle n^\uparrow_i + n^\downarrow_i\rangle$, and 
performing the limit $e^{-U}\rightarrow0$, which yields $\langle n^\uparrow_i n^\downarrow_i\rangle=0$, in the expression for 
the currents (Eq.~\ref{Eq:current}), one can identify the following mean-field current-density relation, already familiar from TASEP,
\begin{equation}
J=\tau\left(1-\tau\right).
\end{equation}
Due to the large potential in this regime, particles on different lanes are not able to `overtake' each other. Hence, the ratio of 
the densities in both channels is fixed to the value given by the reservoir densities at the left boundary,
\begin{equation}
\frac{\ru}{\rd}=\frac{\ru_\text{res}}{\rd_\text{res}}~.
\end{equation}
In Fig.~\ref{fig:dep_U} we have used equal reservoir densities for the upper and the lower lane, and as a consequence  the bulk densities of both lanes are equal.

The exact phase behavior can be determined by relating the boundary conditions of the two-lane system to the corresponding boundary 
conditions of the effective one-lane TASEP, for which the exact phase diagram is known. For reservoir densities 
$\ru_\text{res},~\rd_\text{res}<0.5$, the effective entering rate is obtained by simply adding both reservoir densities:
\begin{eqnarray}
\alpha_\text{eff}&=&\ru_\text{res}+\rd_\text{res}.
\end{eqnarray}
The effective exiting rate displays a more complex dependence on the boundary processes because the individual exiting rates influence 
the exiting  current on both channels.
To find a good estimate of the effective exiting rate, we consider the average time a particle spends on the last lattice site before it 
leaves the channel. This time is the inverse of the corresponding exiting rate. The weight of both waiting times is given by the ratio
of particles in the upper and lower lane. Hence, a fraction $\ru_\text{res}/(\ru_\text{res}+\rd_\text{res})$ of all particles
spend $1/\beta^{\uparrow}$ time units on the last lattice site, and a fraction $\rd_\text{res}/(\ru_\text{res}+\rd_\text{res})$ 
of the particles  $1/\beta^{\downarrow}$ time units.
The average time is the sum of both times weighted with their frequency,
\begin{displaymath}
\frac{\ru_\text{res}}{\ru_\text{res}+\rd_\text{res}}\frac{1}{\beta^{\uparrow}}+\frac{\rd_\text{res}}{\ru_\text{res}+\rd_\text{res}}\frac{1}{\beta^{\downarrow}}\, ,
\end{displaymath}
yielding the effective exiting rate,
\begin{eqnarray}
\beta_\text{eff}&=&\frac{\left(\ru_\text{res}+\rd_\text{res}\right)\beta^\uparrow\beta^\downarrow}{\ru_\text{res}\beta^\downarrow+\rd_\text{res}\beta^\uparrow}.
\label{Eq:beta_eff}
\end{eqnarray}

For reservoir densities $\ru_\text{res},~\rd_\text{res}>0.5$, the double occupation density at the reservoir does not vanish and
is transported into the system.
Hence, the system can exhibit total densities larger than one  if its bulk behavior is determined by the left boundary. In this case, the description 
we introduce below for the regime of intermediate coupling applies.

\section{Current-Density Relation}\label{CDR}

The interaction between neighboring particles directly affects the transport properties of the system. 
The current's dependence on the bulk densities is very sensitive on the coupling. Above a certain value the current-density relation changes 
qualitatively resulting in new phases and phase transitions as we show in the following section.

\subsection{One-Site Cluster Approximation \label{Sec:one-site_cluster}}

With increasing coupling strength $U$, the occupation numbers of the same site on different lanes become more and 
more correlated and a simple mean-field approximation fails. However, by employing a one-site cluster approximation 
we obtain a valuable expression for the currents depending on the bulk densities as demonstrated in~\cite{Peschel}.
To account for correlations between the same site on different lanes, we introduce, besides the single particle 
densities $\ru_i=\langle n_i^\uparrow\rangle,~\rd_i=\langle n_i^\downarrow \rangle$, the double occupation density on 
site $i$, $\kappa_i:=\langle n^\uparrow_i n^\downarrow_i \rangle $ as a new variable~\cite{PhysRevA.45.8358}. 

Then, the probabilities for the other three particle configurations on site $i$, unoccupied or occupied by one particle
either on the upper or the lower lane, can be expressed in terms of $\rho_i^\uparrow,~\rho_i^\downarrow$ and $\kappa_i$.  
We neglect all other correlations and employ the standard decoupling approximation scheme there.

Assuming spatially homogeneous density profiles,  the currents on the upper and the lower 
lane can be expressed in terms of  $\rho^\uparrow,~\rho^\downarrow$ and $\kappa$. The details of the calculations are presented in Appendix~\ref{App:A}. We obtain,
\begin{eqnarray}
j^\uparrow=\ru\left(1-\ru\right)+\mu~, \nonumber \\
j^\downarrow=\rd\left(1-\rd\right)+\mu~,
\label{eq:current}
\end{eqnarray}
where $\rho^\ud(1-\rho^\ud)$ is the particle current known from TASEP and $\mu=\kappa-\ru\rd$ is the correlation correction reducing the current compared to the case without any coupling.
Here, the double occupation density $\kappa$ is the positive solution of  the quadratic equation,
\begin{equation}
0=(1-e^{-U})\kappa^{2}+\left[1-(1-e^{-U})\rho\right]\kappa-e^{-U} \ru \rd~,
\label{Eq:quadr_kappa}
\end{equation}
where $\rho=\ru+\rd$ is the total density. We will employ these results in the following section within 
the framework of an extremal-current principle to investigate the phase behavior as a function of the coupling strength.

In agreement with the considerations in the previous section, the double occupation density vanishes in the 
limit $U\rightarrow\infty$ , while it simplifies to $\kappa=\ru\rd$ for $U\rightarrow0$. The latter limit connects the 
one-site cluster approximation to the simple mean-field approximation which is accurate for small potentials.

\subsection{Dependence of the Currents on the Potential $U$ \label{Sec:J_dep_U}}

\begin{figure}[h]
\centering
\subfigure[~Single-channel current, $j^\uparrow$, for $U=0.3$]{
\includegraphics[width=0.225\textwidth]{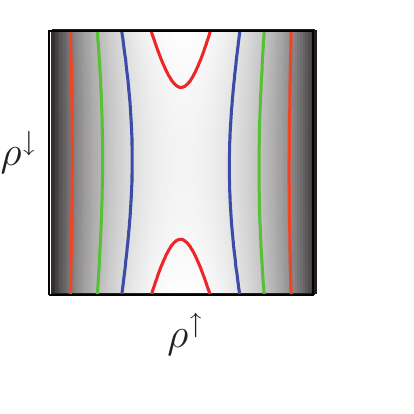}
\label{fig:subfig1}
}
\subfigure[~Single-channel current, $j^\uparrow$, for $U=2$]{
\includegraphics[width=0.225\textwidth]{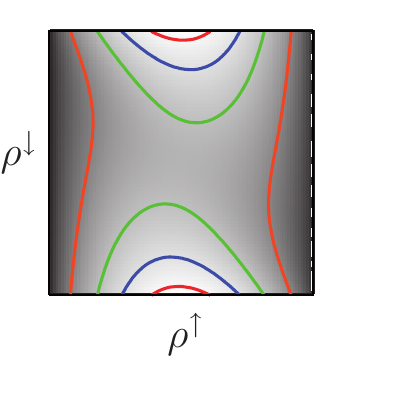}
\label{fig:subfig2}
}
\subfigure[~Total current, $J$,  for $U=0.3$]{
\includegraphics[width=0.225\textwidth]{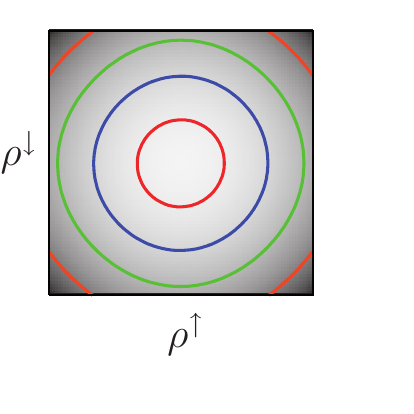}
\label{fig:subfig3}
}
\subfigure[~Total current, $J$, for $U=2$]{
\includegraphics[width=0.225\textwidth]{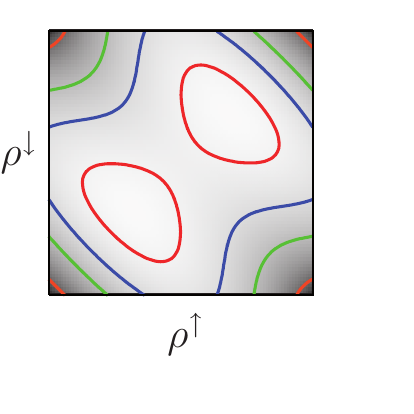}
\label{fig:subfig4}
}
\caption{(color online) Contour plots for  the individual (top) and the total current (bottom), depending on the bulk densities, for $U=0.3$ (left) and $U=2$ (right) using~Eq.~(\ref{eq:current}). On grey scale black corresponds to vanishing current, and white to the respective maximal currents $j^\uparrow_\text{max},~J_\text{max}$. The colored (grey) contour lines indicate currents of value $0.95\cdot j^\uparrow_\text{max}$  and $0.95\cdot J_\text{max}$ (red), $0.8\cdot j^\uparrow_\text{max}$ and $0.8\cdot J_\text{max}$ (blue), $0.5\cdot j^\uparrow_\text{max}$ , and $0.5\cdot J_\text{max} $(green), $0.3\cdot j^\uparrow_\text{max}$ and $0.3\cdot J_\text{max}$ (orange) from inside to outside. Increasing the obstruction strength, the single maximum splits into two, separated by a saddle. The transition happens at a critical value of the interaction, $U_\text{C}=\ln4\approx1.4$.  }
\label{Fig:U03}
\end{figure}
The current displays a sensitive dependence on the strength of interaction, $U$. 
For small interaction strength, the  currents on each lane are almost independent of the density on the other lane. Indeed, 
these currents are approximately parabolic with respect to the density in the respective lane, see Fig.~\ref{fig:subfig1}. 
The maximal current on the upper channel $j^\uparrow_\text{max}$ occurs for $\ru=1/2,~\rd=0$ and $\ru=1/2,~\rd=1$ since
the obstruction does not affect the transport on the upper lane for these densities. With an increase in the potential $U$, 
the particle flux  decreases, in particular, for densities around $\ru=\rd=1/2$, as shown in Fig.~\ref{fig:subfig2}.  

This behavior is also reflected in the total current as shown in Figs.~\ref{fig:subfig3} and~\ref{fig:subfig4}. For small 
potentials, the total current in bulk displays a single maximum located at $\ru=\rd=1/2$, see Fig.~\ref{fig:subfig3}. 
In this regime the potential only affects the value of the maximum but not its position,\emph{ i.e.} it does not change the 
topology of the phase diagrams.  Beyond a critical value of the potential, $U_\text{C}=\ln 4\approx 1.4$, the total current 
displays  a qualitatively different behavior, as has been described in Ref.~\cite{Peschel}. At the critical value, two maxima separated by a saddle evolve in the current-density 
relation for the total current, see Fig.~\ref{fig:subfig4}.  The location of these maxima is evaluated in App.~\ref{App:A}. Upon further increasing $U$, the maxima move apart. 
In the limit 
of large potentials two elongated maxima evolve and the saddle becomes a valley located at $\ru=1-\rd$. The 
bimodal structure leads to a richer phase diagram than in the weak coupling regime, $U<\ln 4$. Similar extrema 
in the currents were found previously for one-channel systems, \emph{e.g.}, when 
next-nearest neighbor interactions are taken into account~\cite{Popkov}.

\subsection{Influence of the Potential $U$ on the Right Boundary}

\begin{figure}[t]
\centering
\includegraphics[width=0.4\textwidth]{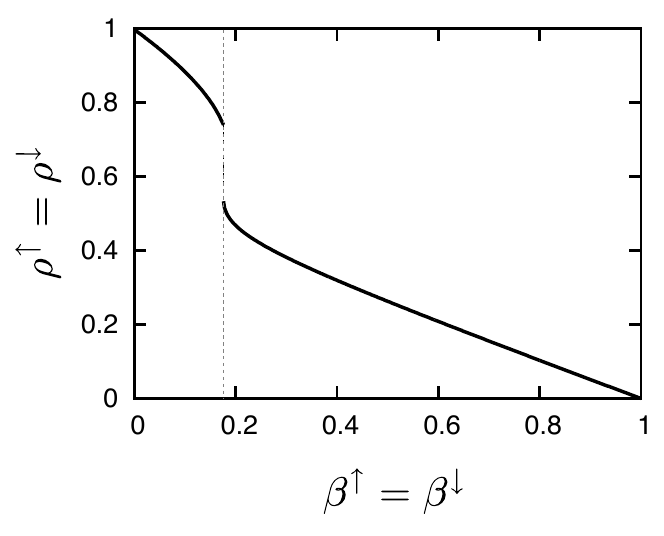}
\caption{The dependence of the bulk densities, $\ru=\rd$, on equal exiting rates $\beta^\uparrow=\beta^\downarrow$ for systems 
determined by the right boundary. It is obtained by evaluating Eq.~(\ref{Eq:ex_current}). The interaction strength  $U=3.0$ 
is above the critical value $U_\text{C}$. 
\label{fig:right_bound}}
\end{figure}
At the right boundary ($i=L$)  exiting rates control the currents out of the system:
\begin{eqnarray}
j^\uparrow_\text{EX}&=&\beta^\uparrow\ru_L,\nonumber \\
j^\downarrow_\text{EX}&=&\beta^\downarrow\rd_L.
\label{Eq:ex_current}
\end{eqnarray}
These currents are either determining the system or are virtual currents which are important for predicting the phase behavior in the system as explained in the following section.
If these exiting currents also set the bulk currents, \emph{i.e.} if $j^\uparrow=j^\uparrow_\text{EX}$ and  
$j^\downarrow=j^\downarrow_\text{EX}$, we can compute the bulk densities (which then equal the densities at the 
right boundary) depending on the exiting rates via Eq.~(\ref{Eq:ex_current}).  The densities at the right boundary 
also play a key role for determining the phase diagrams as we will see below.	
For a small  coupling strength, we find $\rho^\ud\approx 1-\beta^{\uparrow \downarrow}$, as familiar from TASEP. The transport properties 
change rapidly when repulsion between particles increases. In particular, for potentials larger than $U_\text{C}$ the double-maxima structure
of the bulk current comes into play and causes a discontinuous dependence of the bulk density on the exiting rates. Such a jump in the
densities is exemplified in Fig.~\ref{fig:right_bound} for the case of equal exiting rates.

\section{Phase Behavior\label{PB}}

The system's phases depend on the boundary conditions as well as on the strength of internal obstruction. 
In the following, we employ the extremal-current principle to evaluate the analytic expressions for the currents  
obtained in the previous section to get insight into the phase behavior. We then  discuss these phases 
and point out some special features arising from the coupling.

\subsection{Extremal-Current Principle}

The extremal-current principle (ECP) often governs the phase behavior of driven diffusive systems~\cite{Popkov,hager2001min,krug-1991-76}. So far the ECP has only been established for one-dimensional systems. Here we describe a 
 two-dimensional generalization and show that it successfully describes the phase behavior of our two-lane model.  We start with a short description of the standard one-dimensional ECP, and then
extend it to two coupled lanes.

The ECP for transport on a single lane can be formulated by considering two characteristic velocities. 
The first is the collective velocity, $v_c=\partial j/\partial \rho$, 
which reveals information of the stability of a given bulk density $\rho$ against perturbations: only densities with $v_c=0$ (as well as those
determined by the boundaries) are stable. The second quantity is the shock velocity $v_s=\left[j(\rho_1)-j(\rho_2\right]/\left(\rho_1-\rho_2\right)$ that gives
the direction in which a domain wall between two densities, $\rho_1$ and $\rho_2$, travels.  In this way, $v_s$ determines which of  both
densities,  $\rho_1$ or $\rho_2$, dominates. To find the system's bulk density,  it therefore suffices to first identify the stable
densities, using the collective velocity, and then, by pairwise comparison via the shock velocity, single out the bulk density. 
These considerations are summarized by the extremal-current principle:
\begin{eqnarray}
j=\max_{\rho\in[\rho_+, \rho_-]}j(\rho)\ \textrm{for}\ \rho_+>\rho_-,\nonumber \\ 
j=\min_{\rho\in[\rho_+,\rho_-]}j(\rho)\ \textrm{for}\ \rho_+<\rho_-, 
\end{eqnarray}
where $\rho_+$ is the density at the left boundary, and $\rho_-$ is the density at the right boundary.
Hence, the system is either determined by the entering or exiting current or by an extremal current corresponding 
to a density in between the boundary densities.

\begin{figure}[t]
\centering
\includegraphics[width=0.4\textwidth]{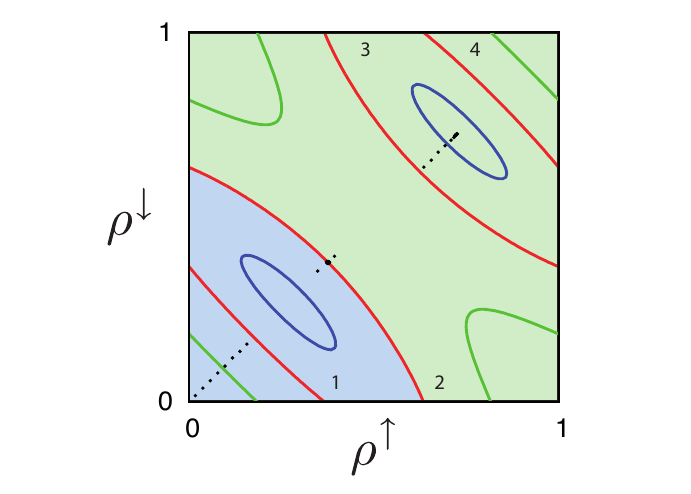}
\caption{(color online) Contour plot of the total current, depending on the bulk densities.
The bulk densities emerging for special values of equal reservoir densities ($\ru_\text{res}=\rd_\text{res}$), namely
continuously increasing from $0$ to $1$, are displayed as black dots.  
The other parameters are  $\beta^\uparrow=\beta^\downarrow=0.3$ and $U=3.0$. The red contour
line corresponds to the total exiting current as emerges if the right boundary determines the bulk densities. 
It marks the transition from the minimal (blue area, lower left corner) to the maximal (green area, upper right corner) current principle and two phase transitions. 
In the lower left corner the system is in the IN/IN phase, crossing the red contour line it enters the EX/EX phase. Upon further increase in the reservoir densities  the IN/IN phase is reached again in the upper right corner, before 
the MC/MC phase arises where the densities are limited by the bulk properties.}
\label{U3xx0303}
\end{figure}

On two coupled lanes, the currents in bulk are generically influenced by both lanes.  We therefore have to consider the dependence of the
currents on both $\ru$ and $\rd$. 
As in the one-dimensional case, either the maximal or the minimal (total) current (see Fig.~\ref{U3xx0303} blue and green area) determines the transport in the system, and the velocities $v_c$
and $v_s$ govern which of both scenarios is realized. However, in order  to decide which of both cases applies it is no longer sufficient
to compare the densities at the boundaries. Because of many potentially conflicting cases a rigorous derivation of the ECP provides a considerable challenge. We have, however, observed that the following intuitive version of the ECP describes our model's phase behavior in the full parameter space.

In the one-dimensional ECP an extremal current belonging to a density in the interval determines the system.
In the two-dimensional scenario the interval is replaced by an rectangle bounded by the boundary densities, $(\rho_L^\uparrow,\rho_L^\downarrow)$ and $(\ru_\text{res},\rd_\text{res})$. Depending on the boundary conditions either the minimal or the maximal current gives the bulk currents. The currents, which have to be considered are the entering and the exiting current or a mixture of both, i.e. one lane is determined by the left while the other one is determined by the right boundary. The exiting current can be calculated by equating Eqs.~(\ref{eq:current}) and (\ref{Eq:ex_current}) with \eqref{Eq:quadr_kappa}. For the specific example $U=3,~\beta^\uparrow,~\beta^\downarrow$, the corresponding contour line of the current-density relation consists of several disjoint lines, see Fig.~\ref{U3xx0303} where the red line has four parts which are denoted as 1,2,3 and 4. Then, the part that includes the point $(\rho_L^\uparrow,\rho_L^\downarrow)$ marks the
boundary where the maximal or minimal current is selected. In the example discussed here this boundary is given by line 2 in Fig.~\ref{U3xx0303}. The extrema to be considered  for the ECP are located at the boundary of the rectangle given by $(\rho_L^\uparrow,\rho_L^\downarrow)$ $(\ru_\text{res},\rd_\text{res})$, or at the extremum on the 
rectangle. This extremum can be inside the rectangle or also at its boundary.

\begin{figure}[b]
\begin{center}
\includegraphics[width=0.45\textwidth]{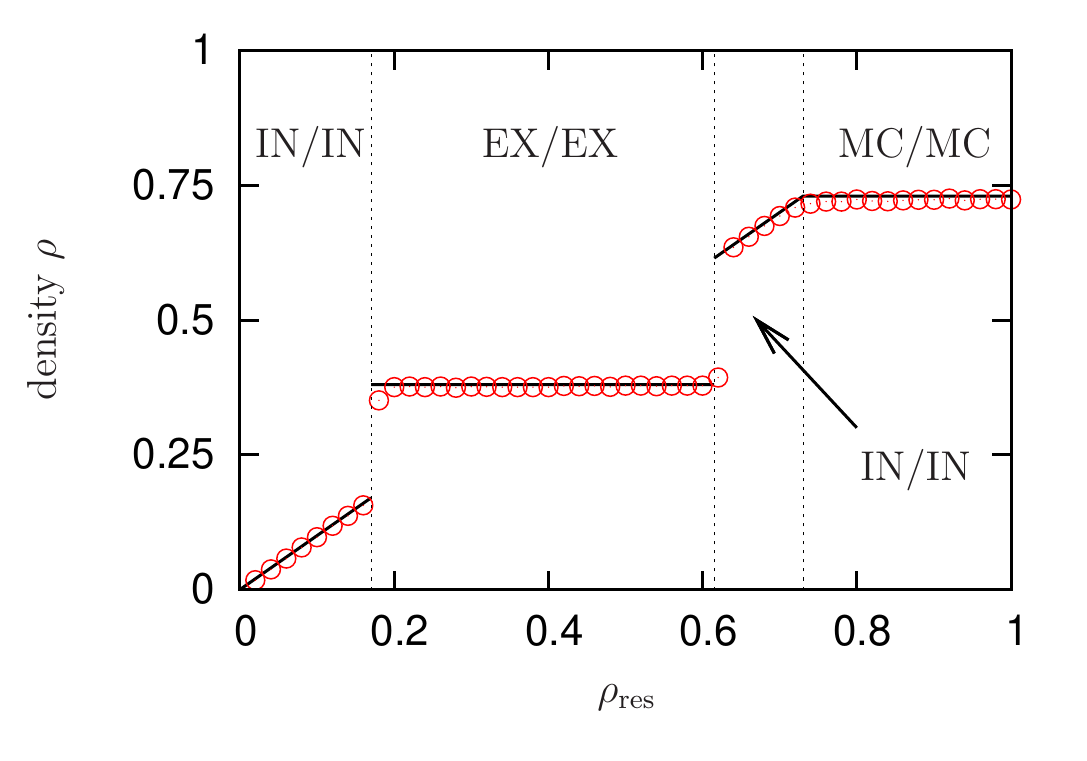}
\caption{(color online) Average bulk densities for increasing reservoir densities $\rho_\text{res}=\ru_\text{res}=\rd_\text{res}$. 
The parameters are set to $\beta^\uparrow=\beta^\downarrow=0.3$ and $U=3$; the situation  corresponds to Fig.~\ref{U3xx0303}. 
The phases introduced here are discussed in Sec.~\ref{Sec:phases}.  Red (grey) circles denote simulation results; black lines correspond to analytical predictions from the ECP.\label{Fig:Densxx0303}}
\end{center}
\end{figure}

To illustrate the extremal-current principle we consider a path where the reservoir densities are gradually increased along the diagonal  
$\rho_\text{res.}=\ru_\text{res}=\rd_\text{res}$ for fixed exiting rates and interaction strength.  Even though we choose this path as an example, our results hold for arbitrary boundary conditions as exemplified in the following.
Fig.~\ref{Fig:Densxx0303} displays the bulk densities  and the phase transitions that occur. 
For small reservoir densities the minimal current is selected, which is given by the  left reservoirs there. Upon crossing the current contour line (1) for the first time, the left reservoir current is larger than the exiting current, and  transport is determined by the exiting rates. Crossing the
contour line that includes the point  $(\rho_L^\uparrow,\rho_L^\downarrow)$ (2), does not give rise to a phase transition, since now the maximum current 
determines the bulk current. In this domain the exiting current is larger than the left reservoir current.  The next
phase transition happens when the  current at the left boundary exceeds the exiting current. Again, the
phase transition occurs at a segment of the red contour line (line 3). Upon a further increase, the second maximum of the current-density
relation is reached, and the maximal-current phase is entered. For reservoir densities larger than $(\ru_{I},\rd_{I})$, \emph{i.e.} for the maximum bulk
current, see Eq.~(\ref{Eq:Max}) in the Appendix, a maximum current is attained.

\subsection{Phases\label{Sec:phases}}

As discussed above the system either adopts its minimal or maximal current, depending on the boundary conditions. These extremal
currents can be either given by one boundary or by an extremum of the current-density relation itself. Hence, we can  distinguish
two classes of phases in the system, boundary- and bulk-induced phases. The first one is highly sensitive to small changes in the
boundary conditions, while in the latter one the densities are determined by the bulk properties and do not depend on the entering and exiting parameters.

\subsubsection{Boundary-induced phases}

The boundary-induced phases depend either on the entering or exiting processes, and  we consequently differentiate between
IN and EX phases. In our model, we employ particle reservoirs at the left boundary, but exiting rates at the right one. 
As a consequence the left and the right boundary influence the bulk densities in qualitatively different ways. 
Indeed, in the case
where both lanes are determined be the left boundary (IN/IN phase), the bulk densities are given by the
reservoirs densities, $\ru=\ru_\text{res},~\rd=\rd_\text{res}$. In contrast, if a system is in the EX/EX phase, only
the total current is fixed to the value given by the right boundary, whereas the bulk densities also depend on the
reservoir densities at the left boundary, for $\ru_\text{res}\lessgtr\rd_\text{res}$ holds $\ru\lesseqgtr\rd$. Further mixed phases (IN/EX or EX/IN) 
may become relevant 
where the current on one lane is determined by the left boundary and the current on the other is fixed by the exiting current.  

\begin{figure}
\includegraphics[width=0.9\columnwidth]{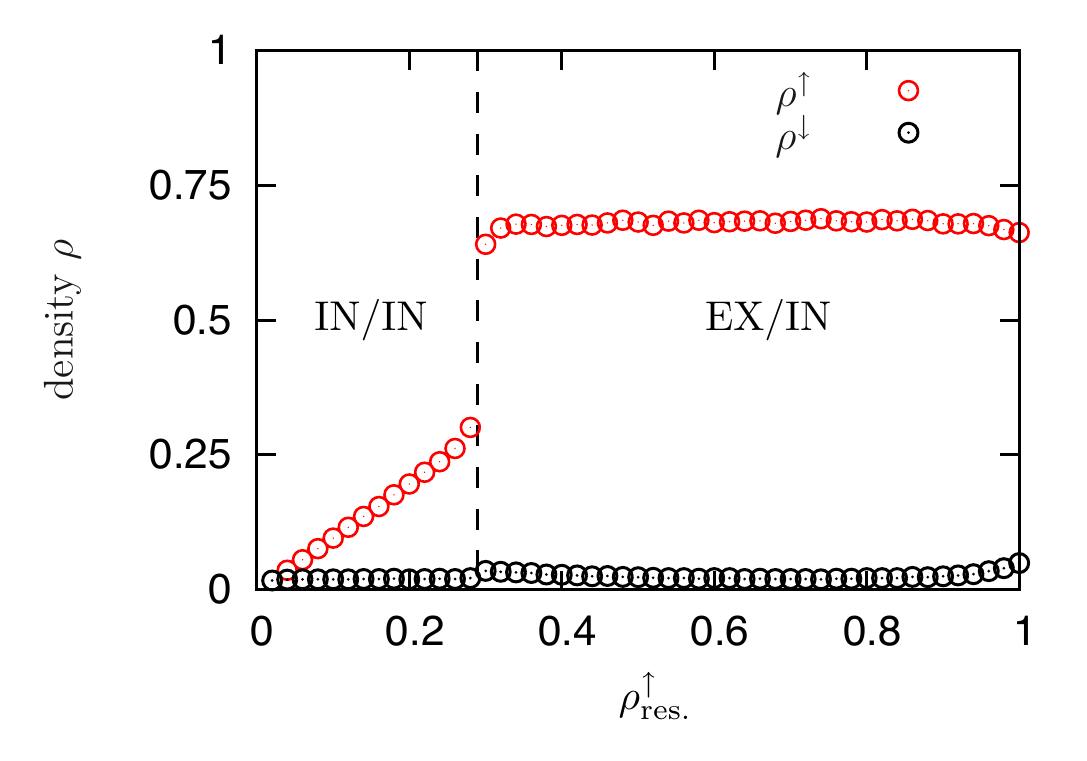}
 \caption{(color online) Average bulk densities for different upper reservoir densities $\rho^\uparrow_\text{res.}$. Parameters are $\rd_\text{res}=0.02,~\beta^\uparrow=\beta^\downarrow=0.3$,  $U=3.0$. If both reservoir densities are small, the system is in the IN/IN phase where both bulk densities are given by the reservoirs. This changes for larger $\ru_\text{res.}$. While the lower lane is still determined by the left boundary, the upper lane is governed by the right boundary (EX/IN phase).   \label{Fig:EXIN}}
\end{figure}

The IN/IN and the EX/IN phase are exemplified in Fig.~\ref{Fig:EXIN} where we show the average bulk densities depending on the upper reservoir density $\rho^\uparrow_\text{res.}$. Parameters are chosen such that a transition from the IN/IN to the EX/IN phase emerges at a certain value of $\rho^\uparrow_\text{res.}$. Only in the IN/IN phase do the bulk densities vary when changing the upper reservoir density $\ru_\text{res.}$; in the EX/IN phase they are almost undisturbed by changes in $\ru_\text{res.}$.

\subsubsection{Bulk-induced phases} 

In the bulk-determined phases, the current is given by an extremum of the current-density relation. 
Here we restrict the discussion  to  the
case where a maximum occurs (MC phase), yet one can find parameter regions where a saddle fixes the phase behavior~\cite{Dipl-arbeit}.

\begin{figure}[t]
\begin{center}
\includegraphics[width=0.45\textwidth]{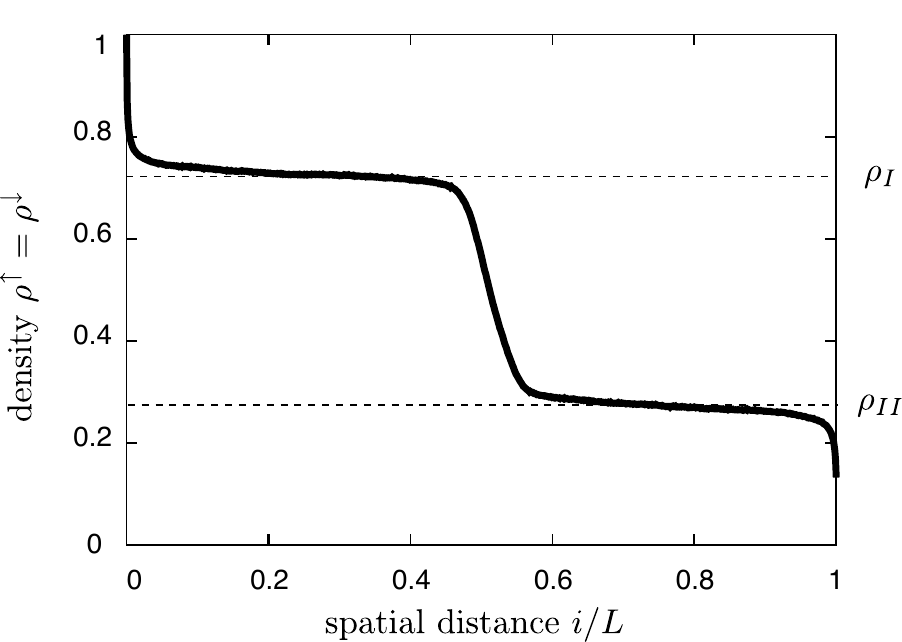}
\caption{Density profile of a state exhibiting maximal current. The 
parameters are $\ru_\text{res}=\rd_\text{res}=1,~\beta^\uparrow=\beta^\downarrow=1$,  $U=3.0$, and $L=1000$ \label{Fig:MC}. The current
remains spatially constant while the density shows a high value at the left and a low value at the right. Both densities correspond
to  maxima in the current-density relation, $\rho_I,~\rho_{II}$.  An unusual localized domain wall forms between them, see text.}
\end{center}
\end{figure}

In the maximum-current phase a localized domain wall can emerge, separating a high-density regime at the left and a low-density regime at the right, 
see Fig.~\ref{Fig:MC}.
This scenario, previously found in Ref.~\cite{krug-1991-76}, occurs if both maxima $(\ru_I,\rd_I)$ and $(\ru_{II},\rd_{II})$, Eq.~(\ref{Eq:Max}) in the Appendix, 
are accessible. Each domain along the lane then corresponds to one of the maximal currents. In particular the current is continuous at the domain wall. 
In Fig.~\ref{Fig:MC}, the reservoir density at the left boundary
is larger than the density corresponding to the second maximum, while the density at the right boundary is smaller than
the density corresponding to the first maximum. Hence, in the vicinity of each boundary the density which is closer
to the respective boundary density arises. Depending on the exact values of the boundary densities, the domain wall
is located between the left hand side and the middle of the system. Increasing the reservoir densities the domain wall moves further to the right.   In Fig.~\ref{Fig:Densxx0303}, the MC/MC phase is shown for symmetric boundary conditions. The black line denotes  the analytically obtained density corresponding to the second maximum of the current density relation. It is in excellent agreement with the simulation results.

\subsection{Phase Diagrams\label{phase_diagrams}}

Employing the extremal-current principle, the phase diagrams can be characterized completely also for unequal boundary densities.. As described above,  there exist several
phases which are either determined by the boundaries or by the bulk transport properties of the system. According to the
ECP the phase transition lines are given by equating the entering  and exiting currents 
or by the structure of the current-density
relation.  The bulk current dominates if the maximum on the rectangle bounded by the corners ($\ru_\text{res},\rd_\text{res}$), ($\ru_{L},\rd_{L}$) 
is not given by either of these points. It is clear that a transition from a left-reservoir  or right-exiting-currents dominated phase  
to a maximum-current phase can emerge only
if the extremum traverses  the boundary of the rectangle considered. Further transitions to an IN/EX and EX/IN phase are identified using again the
extremal-current principle employing contour lines corresponding to the total current in the mixed phases. 

\begin{figure}[t]
\begin{center}
\includegraphics[width=0.45\textwidth]{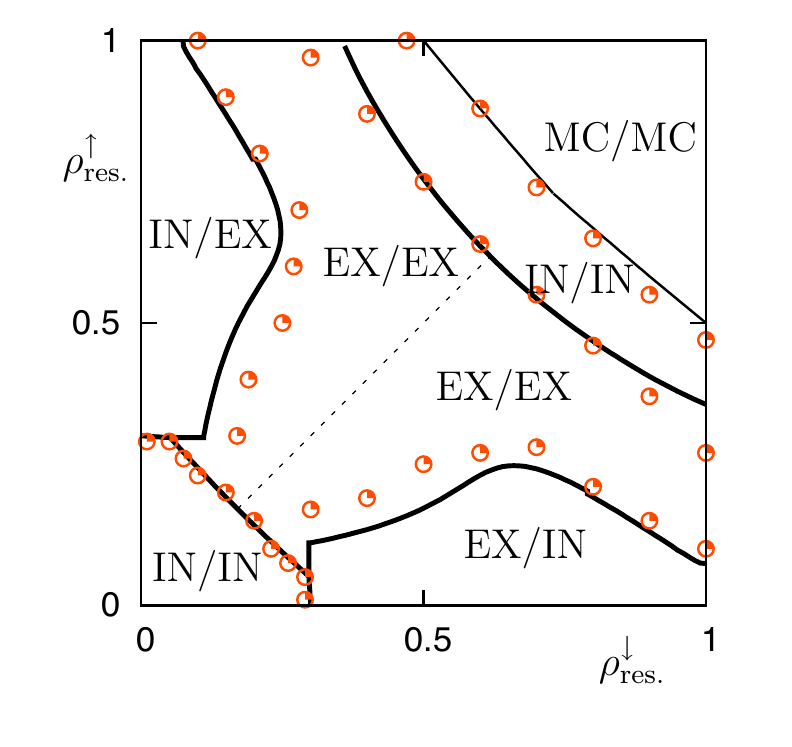}
\caption{(color online) A generic phase diagram with the reservoir densities as control parameters. The exiting rates are
fixed to $\beta^\uparrow=\beta^\downarrow=0.3$ and the interaction strength is set to $U=3.0$\label{Fig:PD}. The red (grey) dots denote simulation results of the transition lines, while the black lines are calculated employing the ECP. Besides
the combinations of phases which are determined by entering or exiting currents,  phases
which do not exist for $U<\ln4$ are also present. Namely a second IN/IN phase, which exhibits a high
density, and the MC/MC, where the maximal current determines the bulk densities, occur.}
\end{center}
\end{figure}
In Fig.~\ref{Fig:PD} we exemplify a phase diagram depending on unequal reservoir densities. 
The exiting rates are fixed to $\beta^\uparrow,~\beta^\downarrow=0.3$ and the potential is set to $U=3$, 
a value where the current-density relation shows two maxima. The parameters are identical to the ones of 
Figs.~\ref{U3xx0303} and \ref{Fig:Densxx0303}. In Fig.~\ref{Fig:PD}, the stochastic simulations (red dots) are in good agreement with the analytic calculations (black lines). Because the latter ones were obtained employing the ECP also for unequal reservoir densities, the strength and generality of the ECP can be confirmed.
Due to the symmetry between both lanes, the phase diagram is
symmetric along its diagonal, $\ru_\text{res}=\rd_\text{res}$.

In the lower left corner, where both reservoir
densities are small, both channels are determined by the entering current (IN/IN phase). 
Increasing only one reservoir
density, the minimal current is no longer given by the entering current on the respective lane. Hence the system 
reaches the EX/IN phase (IN/EX phase). The phase-transition lines between the IN/IN and IN/EX phase, as well as the ones
between the IN/EX  and the EX/EX phase, are given by contour lines of the current on an individual 
lane, Eq.~(\ref{eq:current}). In the middle of the phase diagram, both channels are determined  by the exiting
currents. As mentioned above, only the total current is fixed in this phase, in contrast to the bulk densities which
also depend on the reservoir densities at the left boundary. 
The dotted line marks the region where the bulk densities
on both channels are equal. On this line the bulk densities can be calculated employing Eq.~(\ref{Eq:ex_current}). 
For a further increase in the reservoir densities, the IN/IN phase arises again where the bulk currents as
well as the bulk densities are given by the left boundary.  This phase would not arise if we had 
chosen entering rates, rather than  particle reservoirs,  at the left boundary, because boundary densities larger
than $1/2$ would then not emerge. In our case, such large  densities cause currents larger than the entering current. These dominate the system according to the extremal-current principle. For even larger reservoir densities the crest
of the current-density relation, which marks the phase transition from the IN/IN to the MC/MC phase, is reached, and 
the bulk densities remain constant at the values where the second maximum is located.

\section{Summary and Conclusion\label{summary}}

In this paper we have examined a driven two-channel model where the motion of particles along both channels is
coupled via a repulsive short-range interaction. The latter causes  intriguing new phenomena and phases.
Varying the strength of particle obstruction, three regimes of qualitatively different behavior evolve. First, 
for small coupling the system approximately acts as two uncoupled lanes, \emph{i.e.} the phases and phase diagram
qualitatively correspond to the ones already known from TASEP. These results can be confirmed by employing a
simple mean-field approach or a one-site cluster approximation. Second, for moderate interaction strength the transport
properties of the system are strongly influenced by the obstruction between neighboring particles. This regime emerges
around a critical interaction strength, $U_\text{C}=\ln4$~\cite{Peschel}, where a single maximum of the current-density relation splits into
two, separated by a saddle. Third, when the obstruction is large, the two coupled lanes effectively behave as a single
one. In this case, we have identified a mapping from the parameters governing entering and exiting processes in our
system to effective rates for a single-lane TASEP. This mapping then allows to carry over known results from  TASEP such as its phase diagram. 
Hence, for different strengths of the obstruction a variety of peculiar phases surface which can be accessed by manipulating the system at the boundary 
only. The boundary-sensitive phases respond gradually upon tuning the left reservoir or the exiting rates, whereas the maximal-current phase is robust    
against such changes. 

In contrast to previous work~\cite{Peschel}, we explicitly investigated the two-channel system with a broken lane symmetry.  We thereby followed the proposition in~\cite{Peschel} that the ECP might be generic for multichannel system. As a key result we indeed derive a suitable generalization of the ECP to two dimensions. We thereby approve that not the densities but the currents govern the transitions between the minimal and the maximal current principle, a distinction that cannot be made within one-dimensional or symmetric situations.  The accuracy with which the system's phase behavior can be predicted with our two-dimensional ECP  is astonishing and suggests it for further applications. In further contrast to~\cite{Peschel}  we specifically investigated the dependence on the potential $U$. For small potentials the system's behavior is not sensitive on whether reservoir or rates are chosen at the boundaries. However, for intermediate and large potentials, reservoirs or rates at the boundaries make a difference. For example, the system does not behave like a one-channel system for large potentials if reservoirs are chosen instead of rates at the right boundary, as has been done in~\cite{Peschel}. Because we employ exiting rates at the right boundary, we observe novel effects such as  a strong impact of the potential $U$ on the exiting current (Fig.~\ref{fig:right_bound}), and two transitions instead of one in the bulk densities upon increasing $U$ (Fig~\ref{fig:dep_U}).

It would be interesting to consider also lane changes (resp. spin flips) of the particles as they proceed along the channel. 
Such events are expected in realistic
applications such as intracellular transport, highway traffic on multiple lanes, or hopping 
transport of electrons through
a chain of quantum dots. The correlations induced by frequent lane changes~\cite{pronina-2004-37,pronina-2006-372}, ignoring mutual obstruction,  
are accurately described by a one-site cluster approximation similar to  the one we employed. 
Yet, for rare lane changes a simple mean-field approximation suffices to  describe the arising localized
domain walls~\cite{reichenbach-2006-97}. The combination of obstruction, lane switching and possibly also defects 
constitutes a promising route to discover 
novel and unexpected collective phenomena in driven transport.

In conclusion, we have  shown that the ECP can be generalized to higher dimensions to serve as an appropriate tool for the investigation of driven multi-channel systems. We therefore think that extremal-current principle is a  promising starting point to achieve a deeper understanding of complex transport phenomena.  

\begin{acknowledgments}
This project is supported by the Deutsche Forschungsgemeinschaft  via the SFB TR12 ``Symmetries and Universalities in Mesoscopic Systems"
and
the  German Excellence Initiative via the program "Nanosystems
Initiative Munich (NIM)''.
T.R.\@ acknowledges funding by the Elite-Netzwerk Bayern.
\end{acknowledgments}

\appendix

\section{Current-Density Relation\label{App:A}}

In this Appendix, we  derive an analytic description for the  current-density relation within a one-site cluster approximation.

The individual currents onto a site $i$ can be obtained be evaluating the particle fluxes onto this site,

\begin{eqnarray}
j^\uparrow_i=&&\Big\langle e^{-U}\left(n^\uparrow_{i-1}-n^\uparrow_{i-1}n^\downarrow_{i-1}\right)
\left( n^\uparrow_{i}-n^\uparrow_i n^\downarrow_i\right)\Big\rangle \nonumber \\&+&\Big\langle
\left(n^\uparrow_{i-1}-n^\uparrow_{i-1}n^\downarrow_{i-1}\right)\left(1-n^\uparrow_{i}-n^\downarrow_{i}+n^\uparrow_{i}n^\downarrow_i\right)\Big\rangle\nonumber \\
&+&\Big\langle n^\uparrow_{i-1}n^\downarrow_{i-1}(1-n^\uparrow_{i})\Big\rangle\,, 
\label{Eq:current}
\end{eqnarray}
and similarly for $j^\downarrow_i$. 
This expression is evaluated employing the one-site cluster approximation discussed in 
Sec.~\ref{Sec:one-site_cluster}.  The essence of the approach consists of considering 
all four possible states of the two opposing sites, whereas all correlations between neighboring sites are factorized.  
Thus, a complete description is achieved in terms 
the mean double-occupation 
density $\kappa_i=\langle n^\uparrow_i n^\downarrow_i\rangle$ besides the average densities
$\ru_i=\langle n_i^\uparrow\rangle$ and $\rd=\langle n_i^\downarrow\rangle$. 
Then the closure relation for the currents is derived
\begin{eqnarray}
j^{\uparrow}_{i} =&& e^{-U}(\rho^\uparrow_{i-1}-\kappa_{i-1})(\rho^\uparrow_{i}-\kappa_{i})\nonumber\\
         &+&(\rho^\uparrow_{i-1}-\kappa_{i-1})(1-\rho^\uparrow_{i}-\rho^{\downarrow}_{i}+\kappa_{i})\nonumber \\
         &+&\kappa_{i-1}(1-\rho^\uparrow_{i}).
\label{Eq:current_site}
\end{eqnarray}

Similar to Eq.~(\ref{Eq:current}),  the time evolution for the  double-occupation density $\kappa_i$  involves averages of products of occupation variables $n_i$. 
Within the same truncation of the hierarchy, one finds,
\begin{align}
\partial_t\kappa_i =& e^{-U}(\ru_{i-1}-\kappa_{i-1})(\rd_i-\kappa_i)\nonumber \\
&+e^{-U}(\rd_{i-1}-\kappa_{i-1})(\ru_i-\kappa_i)\nonumber\\
&+\kappa_{i-1}(\ru_{i}+\rd_{i}-2\kappa_{i})\nonumber\\
&-\kappa_{i}(2-\ru_{i+1}-\rd_{i+1})\,.
\end{align}
In the steady state and for spatially homogeneous density profiles, the double occupation density $\kappa$ 
obeys a quadratic equation with solution,

\begin{align}
\kappa = \frac{\lambda\rho-1+ \sqrt{\left(\lambda\rho-1\right)^2 +4 \lambda (1-\lambda) \ru\rd}}{2 \lambda} \, .
\label{eq:kappa}
\end{align}
Here, $\lambda= 1- e^{-U}$ corresponds to the inverse dwell time in the obstructed case and $\rho = \ru + \rd$ to the total density. Combining this result with the closure relation for the current, Eq.~(\ref{Eq:current_site}), the closed expression for 
the current-density relation follows,
\begin{eqnarray}
j^\uparrow=\ru\left(1-\ru-\rd \right)+\kappa, \nonumber \\
j^\downarrow=\rd\left(1-\ru-\rd \right)+\kappa\,.
\end{eqnarray}

The total current $J=j^\uparrow+j^\downarrow$ displays a single a maximum for small interactions $U$ 
located at $\ru= \rd = 1/2$. This maximum is replaced a saddle for strong coupling and two maxima of equal height appear on the diagonal. 
These new maxima are located at,
\begin{equation}
\ru_{I,II}=\rd_{I,II}=\frac{1}{2}\pm\frac{\sqrt{1-5e^{-U}+4e^{-2U}}}{4(1-e^{-U})}\,.
\label{Eq:Max}
\end{equation}
These solutions are only real for potentials larger than the critical value, $U_\text{C} = \ln 4$.

\end{document}